# Crossover between Rayleigh-Taylor Instability and turbulent cascading atomization mechanism in the bag-breakup regime

Nicolas Rimbert, Guillaume Castanet
Nancy University
LEMTA, ESSTIN, 2 av. de la Forêt de Haye,
F-54504 Vandoeuvre-lès-Nancy cedex

**Abstract** The question whether liquid atomization (or pulverization) resorts to instability dynamics (through refinements of Rayleigh-Plateau, Rayleigh-Taylor or Kelvin-Helmholtz mechanism) or to turbulent cascades similar to Richardson and Kolmogorov first ideas seems to be still open. In this paper, we report experimental evidences that both mechanisms are needed to explain the spray drop PDF obtained from an industrial nozzle. Instability of Rayleigh-Taylor kind governs the size of the largest droplets while the smallest ones obey a PDF given by a turbulent cascading mechanism resulting in a log-Lévy stable law of stability parameter close to 1.68. This value, very close to the inverse of the Flory exponent, can be related to a recent model for intermittency modeling stemming from self-avoiding random vortex stretching.

## 1 Introduction

Lognormal probability density functions (PDF) are widely used by experimental analysts in many fields, including liquid pulverization. To explain their widespread appearance in many fields, Kolmogorov indirectly pioneered turbulent atomization modeling [1] by devising a discrete Markov process converging toward a lognormal PDF. Later Obukhov [2] used this result to describe the statistical distribution of intermittent dissipation in turbulent flows and this has later been retained by Kolmogorov [3] in his famous K62 modeling of turbulence intermittencies. Many works have been since done in this field and Frisch's book [4] is an excellent review. Maybe the most influential scientist still to be cited is Mandelbrot [5] who coined the word fractal, developed the multifractal formalism and made many contributions to economics where he widely used log-stable distribution (for a definition of Lévy stable laws many textbooks do exist now but [6] is still a good reference). Unfortunately, concerning turbulence or atomization modeling, Mandelbrot merely developed general ideas but no reproducible laws. While Schertzer and Lovejoy [7] emphasized their role in geophysics as universal multifractals (thanks to a generalized central limit theorem), Kida [8,9] explained empirically the statistical laws of turbulence intermittencies with a log-stable law of stability parameter 1.65. Though this is still debated nowadays[10], a recent advance in this field can be found in [11,12] where Kida's results are discussed and proved thanks to a self-avoiding random vortex stretching process. In this modeling, the topological constraint of non intersection, applied to a vortex tube, enforces the value of the stability parameter to be the inverse of Flory's exponent (i.e. 1/.588 or 1.70), a scaling exponent well known in polymer physics [13]. The stable law is proved to be fully asymmetric to the left and its scale parameter can be related to the important scale of turbulent flows: Kolmogorov's and Taylor's scales.

As for turbulent atomization modeling, few improvements have been made in this field since Kolmogorov's work: while experimentalists still resort to a variety of empirical laws [14], some of them close to the lognormal law (such as the upper limit lognormal Evans law or the log-Weibull law), theoreticians widened their view to either log-infinitely divisible distributions [15], multifractal analysis [16], or refinements of Kolmogorov first modeling [17]; all of which not very helpful for the experimentalists. In a more practical way, in [18], it has been shown that log-stable laws (which now can be easily computed thanks to FFT) are also good candidates to model some turbulent spray PDF. Atomization being a growing field of interest a more detailed review can be found for instance in [19,20].

However, though multistep cascading mechanisms are very useful to describe turbulent intermittencies, it is appropriate to question whether they pertain to atomization modeling: there are numerous records of primary atomization and secondary atomization, but, except in the so-called catastrophic breakup [21], almost never of ternary atomization (i.e. no third steps!). This may actually be related to the fact that after two breakup events, some air is entrapped in the neighborhood of the drops in what is often called the "added mass" so that liquid droplets are no longer sheared. Moreover recent evidence indicates that a cascading mechanism seems to be still appropriate [22] for high-speed sprays. For lower speed, most analysis made in the non turbulent regime, resort to instability theory. It usually leads to a competition between surface tension effect (Rayleigh-Plateau mechanism), acceleration of a droplet in the ambient air (Rayleigh-Taylor mechanism) and shear instability appearing on the edge of the droplet (Kelvin-Helmholtz mechanism, for a review of these mechanisms, see [23] for instance). In a recent study of the so-called non turbulent bag breakup regime [24], the Rayleigh-Taylor mechanism is widely used to explain experimental results.

In this work, we will show firstly, that for an industrial nozzle, in the bag-breakup regime, Rayleigh-Taylor instability can explain the first stage of the breakup (as well as accepted values of the so-called bag breakup regime) It is then reported how a turbulent cascade mechanism seems to be necessary to describe the finer droplets PDF resulting from the burst of the bag. In a way, this can be related to the late stage of the turbulent mixing in the Rayleigh-Taylor instability where recently a Kolmorogov-like cascading scenario has been put in evidence (cf. [25] for a description of the one-phase RT instability developing between a hot gas and a cold gas).

## 2    *Experimental setup*

Experimental setup was devised to test several kinds of industrial nozzles, among which two Lechler Nozzles (ref. 665-042 and 665-122) were tested. Their aperture is depicted in Fig. 1: it is made of two identical circle segments pieced together. The measured value of parameter 2*a* and 2*b* of Fig. 1 are respectively 977 $\mu m$ and 646 $\mu m$ for ref. 665-042. and 1185 and 864 $\mu m$ for ref. 665-122. Their equivalent radiuses $r_{eq}$ are thus respectively 452 $\mu m$ and 488 $\mu m$. Results obtained from either nozzle are thus quite similar. Their nominal flow rate is around 80 liters of water per minutes (varying with pressure). Several pressures were tested but most measurements were made at 8 bars and 15 bars.

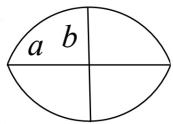
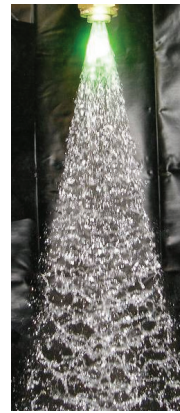

**FIG. 1: sketch of the nozzle orifice and photograph of the experimental set-up (Lechler Nozzle ref. 665-042, 8 bars)**

For the 665-122 nozzle, for a given pressure of 15 bars, the liquid velocity $U$ was found to be 41 m/s and the standard deviation $u'$ around this velocity was 2 m/s (cf. FIG. 3.). Let us notice that these velocities are independent of the droplets size. Data were collected using a Dantec Dynamics PDA (Phase Doppler Anemometer) and a green argon continuous laser (wavelength: 514.5 nm). The PDA equipped with a classic receiver, has been used in refraction mode with a diffusion angle of 72°. According to the manufacturer droplets size can be measured with confidence in a dynamic interval ranging from 1x to 40x (or 1.6 decades). This means that bigger droplets have a tendency to saturate the photomultipliers while smaller ones may not trigger it. However data were collected over three decades. While droplet frequency measurement may not be very accurate on this wide range, caution was made to configure measurements so that small drops resulting from the bag-breakup were accurately resolved (i.e. no aperture mask and a high enough voltage amplification ratio was used in the photomultiplier). Actually the measuring volume was an ellipsoid of dimension *600μm* x *600μm* x *4000μm* so that only droplets of diameter inferior to *600 μm* could be fully contained in the measuring volume. This setup resulted in a detected maximum droplet size of 1754 *μm*. Moreover droplets bigger than *600μm* were mostly detected when they cut the volume measure on the side of it, i.e. when the amount of light they emitted was small enough not to saturate the photomultipliers. This can be seen by comparing the measured transit time to an idealized transit time given by:

$$TT_{ideal} = \frac{(600).10^{-6}}{U} \quad (1)$$

Here the Gaussian characteristic of the laser beam is neglected and the droplets are considered as point particle. Results of such a comparison can be found in FIG. 2 which depicts the evolution of the ratio of the measured transit time (given by the PDA) to the idealized transit time as a function of droplets' diameters. It can be seen that larger droplets have a very short measured transit time indicating that they were mainly cutting the measuring volume on its borders. Accordingly, their number cannot be assessed with certainty. Following, this analysis, the center of the measurement range where the statistics of the droplets is adequately reported can be estimated to be around *50μm* resulting in a PDF fully resolved somewhere between 10 and *400μm* (but this is an harsh estimate). Note finally that since the three detectors are in line, what is truly measured by the PDA, are two inline radii of curvature of the interface between water and air; when these radii share a common value, the latter is then assimilated to the radius of the measured drop. This (common) approximation will be naturally made in this paper.

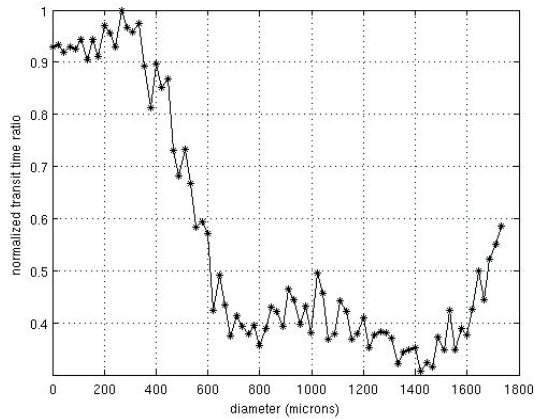

**FIG. 2: comparison between the measured transit time and the ideal transit time of formula (1) showing that big droplets are mostly detected on the edge of the measurement volume. The overall intensity**

## 3    *Experimental results*

As PDA collects both size and velocity of droplets, results can be given in the form of joint PDF. Figure 3 depicts such kind of representation. Middle picture is the joint velocity-diameter PDF and lower picture is the joint velocity-magnitude PDF (magnitude is here defined as the decimal logarithm of the diameter; origin is chosen such that magnitude 0 stands for 1 $\mu m$).

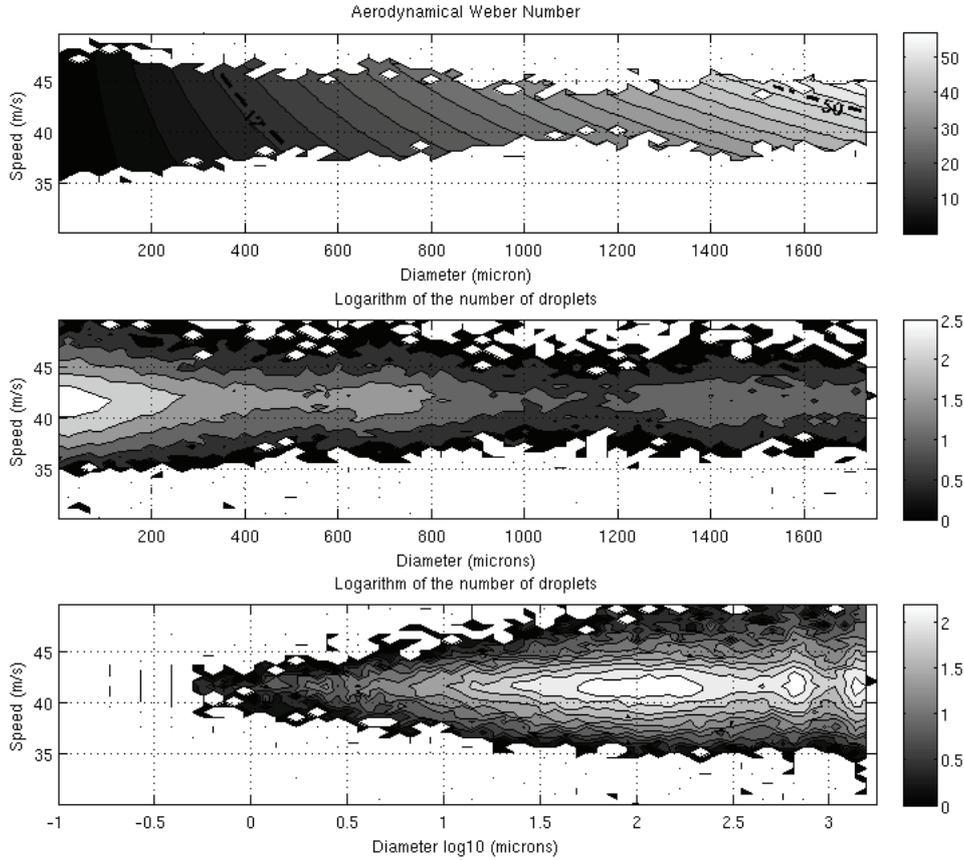

**FIG. 3:** experimental size-velocity PDF. (Lechler Nozzle ref. 665-122, 15 bars, 4cm below the exit) On top the domain of the bag-breakup (12<We<50) is delimited. Comparison between the frequency distribution of the droplet velocity and diameter (in the middle) and the distribution of the velocity and magnitude ($\log_{10}(d)$, on the bottom). It can be seen that a large amount of droplets are located under 100$\mu$m, that is in the aerodynamic stability domain.

The non dimensional parameters governing the stability of a droplet in an air stream are the (aerodynamic) Weber number and the Ohnesorge number defined by:

$$We_{aero} = \frac{\rho_G U^2 d}{\gamma} \quad \text{and} \quad Oh = \frac{\mu_L}{\sqrt{\gamma \rho_L d}} \qquad (2)$$

where $\rho_G$ and $\rho_L$ stand for, respectively, air (gas) density and water (liquid) density; $\gamma$ is the air-water surface tension, $\mu_L$ is the liquid dynamic viscosity while $d$ stands for the droplet diameter. According to [26] at low Ohnesorge number (i.e. mainly for liquids with low viscosity), a Weber number inferior to 12 indicates that the droplet is aerodynamically stable whereas a Weber number located in the interval [12,50] indicates that the droplet shall break in what is called a bag-breakup (cf. Fig. 4). On top of figure 3, iso-contour of Weber number are given in the velocity-diameter plane. From it, it can be seen that most droplets on the millimeter scale are located in the bag-breakup regime whereas droplet whose radii are located under 350 $\mu m$ are aerodynamically stable. Actually the fact that droplets of every size are flowing at the same speed and that the all droplets size-velocity PDF have roughly the same shape on the vertical axis seems to indicate that apart from initial bag breakups, droplets are subsequently not submitted to any particular shear from the ambient air (the opposite would result in smaller droplets going slower, being more decelerated). This suggests that the air surrounding the spray is entrapped in the water flow. This induced a slight wind in the surrounding which could actually be felt during experimental work. Therefore values of the Weber number given by (2) is an overestimation since the velocity of water $U$ shall be replaced by the drift velocity $U_{dr}$ between water and air. If the preceding remark is true this means that when not submitted to primary or secondary breakup, droplets of any size can be stable in the air flow. This may also explain the observed stability of the PDF on the vertical axis.

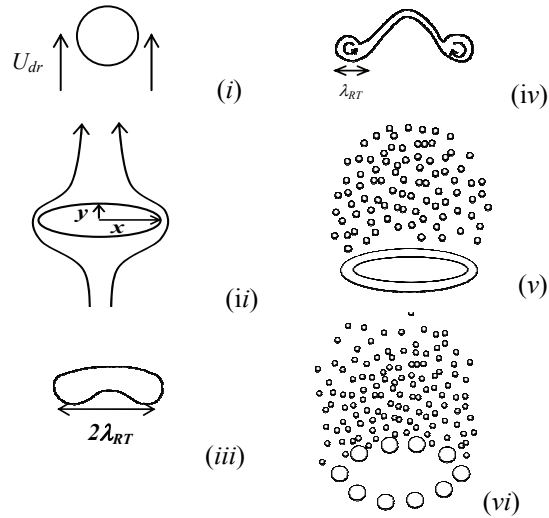

**FIG. 4: the six stages of the bag breakup (cf. [23]): (*i*) the droplets result from the Rayleigh-Plateau mechanism, (*ii*) they elongate in the relative air flow, (*iii*) this results in a Rayleigh-Taylor wave and (*iv*) in the formation of a ring and a bag, (*v*) the blow up of the bag and (*vi*) of the ring leads to a bimodal distribution.**

In these PDF, three peaks do appear (especially when considering the magnitude scale) each corresponding to a peculiar physical mechanism. Position of the three peaks in the distribution are found to be 1355 $\mu m$ (magnitude 3.13), 661 $\mu m$ (magnitude 2,82) and roughly 100 $\mu m$ (and more precisely 200 $\mu m$ considering the Sauter Mean Diameter of the bag). Statistics were made over 50 000 droplets. Since all droplets have the same velocity, this number was large enough to obtain converged results. Note that sizing techniques based on image analysis (usually obtained by high-speed imaging) would lead to a very tedious analysis protocol to obtain such a high number (moreover for a smaller dynamic range).

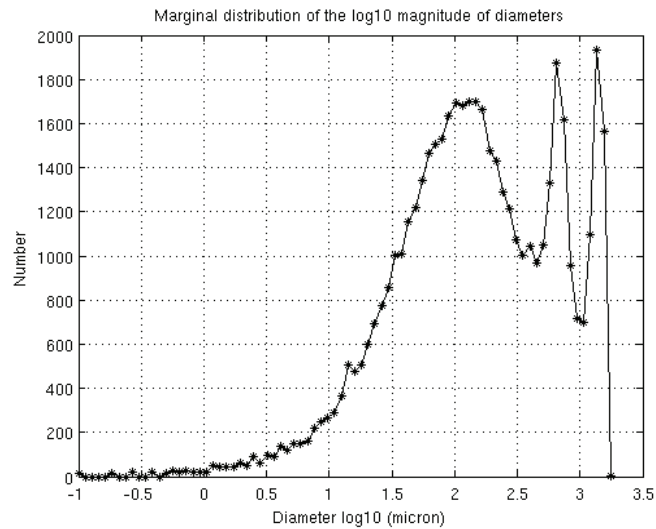

**FIG. 5: marginal size distribution. (Lechler Nozzle ref. 665-122, 15 bars, 4cm below the exit) From right to left, the first droplet is given by the Rayleigh-Plateau theory, the first peak is given by the Rayleigh-Taylor theory while the second peak is another harmonic Rayleigh-Taylor peak. The third peak is the bag breakup peak.**

Figure 4 depicts the classical way of describing the bag-breakup mechanism (cf. [19] for more details). Five stages are usually described: the drop is first deformed by the air stream, this leads to the onset of a Rayleigh-Taylor instability which leads to the formation of a rim and a bag. The bag bursts before the rim which follows soon afterwards (characteristic times can be found in [26,27,28]). Figure 5 shows the marginal magnitude PDF obtained from the joint velocity magnitude PDF. Small size PDF are very similar whereas large size peaks are located at roughly the same place but differs in intensity. Since they mainly corresponds to off-center droplets there is clearly a reproducibility issue concerning their intensity. Since air is entrapped in the water curtain, it can be supposed that the first peak corresponds to unburst droplets while the second peak corresponds to droplets resulting from the fragmentation of the rim and the third peak to the cloud of droplets created by the bursting of the bag. This can be confirmed by the fact that the Sauter Mean Diameter of the bag droplets has been found to be 200 $\mu m$ so that the ratio:

$$\frac{SMD_{bag}}{d_{init}} = \frac{200}{1350} \approx 0.148 \qquad (3)$$

is also very close to the value 0.14 reported in [19] and [21].

## 4   Rayleigh-Taylor breakup

An explanation of the ratio (close to 2) between the diameter of the mother droplet and the diameter of the rim droplets can be developed as in [19] by adapting Rayleigh-Taylor theory. This had been attempted by Kytschtka et al. [29] but without taking into account the deformation of the droplet. Influence of the deformation on the trajectory and drag coefficient of the droplet has been studied in [30, 31] but without really considering the breaking mechanism. This leads to an ordinary differential equation (ODE) of evolution for the deformation, coupled to another ODE for the droplet position. The resulting model is named droplets deformation and breakup (or DDB). The breakup is then assumed to occur naturally at the end of the deformation phase. This hypothesis will be followed hereafter and it will be supposed that the droplet has already deformed into an oblate ellipsoid of major semiaxe $x$ and minor semiaxe y (cf. Fig. 4) when the Rayleigh-Taylor induced breakup starts. Now, let us suppose that $x \ll y$ so that the final shape of the droplet can be assimilated to a disk. If the droplet's Reynolds number is high enough ($Re$ >1000), its drag coefficient $C_d$ can be supposed constant and is given by $C_d$ ranging from 1.7 for incompressible flow to up to 3 for compressible flow [21, 31], 1.7 being very close to the 1.5 high Reynolds drag coefficient of a disc in incompressible flow. Now, let us compute the condition upon which a Rayleigh Taylor wave of wavelength equal to the major semiaxe can grow on the drop surface (cutting the droplet in its center). The fastest growing wavelength is given by:

$$\lambda_{RT,\max} = 2\pi \sqrt{\frac{3\gamma}{f\Delta\rho}} \qquad (4)$$

where $f$ is the deceleration of the disk in the air stream, which is given by:

$$f = \frac{3}{8}\frac{\rho_G}{\rho_L}\frac{x^2}{r^3}C_d U^2 \qquad (5)$$

So that

$$\frac{\lambda_{RT,\max}}{r} = 2\pi \left(\frac{r}{x}\right)\sqrt{\frac{8\gamma}{C_d \rho_G U^2 r}} \qquad (6)$$

The condition $x = \lambda_{RT,\max}$ leads to

$$\left(\frac{x}{r}\right)^4 = 32\pi^2 \frac{\gamma}{\rho_G C_d U^2 r} \simeq 10.8 \qquad (7)$$

where $\rho_G$ = 1.3 kg/m$^3$, $r$ = 675 $\mu m$, $U$ = 40 m/s, $C_d$ = 1.5 and $\gamma$ = 0.072 N.m$^{-1}$. Note that $Re = Ud/\nu_G = 1800$. This deformation corresponds to $x = 1.8r$ which means that when the droplet has reached this deformation, deceleration is high enough for a Rayleigh-Taylor wave of fastest growth to cut the drop in half. According to DDB theory, the maximum deformation is given for this Weber number by:

$$\frac{x}{r} = \frac{We}{6\pi} = \frac{\rho U^2 d}{6\pi\gamma} \simeq 2 \qquad (8)$$

So the value 1.8 can be considered as a reachable state.
Equation (8) also reads

$$We = \frac{64\pi^2}{10.8 C_d} = 38 \qquad (9)$$

This is a hint that by slightly modifying the preceding reasoning more general conclusions can be obtained. Let us set the deformation criteria for the bag breakup to be $x/r = 2 = \lambda/r$ and let us use the experimental $C_d$ = 1.7 [21], one gets

$$\left(\frac{x}{r}\right)^4 = 2^4 = 16 = 64\pi^2 \frac{1}{C_d We} \qquad (10)$$

and finally

$$We_{min} = \frac{4\pi^2}{C_d} \approx 23.2 \qquad (11)$$

which corresponds to Hinze's measurement for free falling drops [14] (the use of $C_d$ =3 leads to $We_{min}$ = 13 which is Hinze's measurement in shock tube experiment). It is also very easy using this simple modeling to evaluate the ratio of the ring-to-bag volume: on Fig. 4 (iii) it can be seen that the droplet is separated in three parts by the two inflexion points of the wavy surface , two third of the drop leading to the rim and one third to the bag. Actual measurement of the rim to mother droplet volume [31] led to the value 70% very close to the proposed 66%. Note that the condition $2x = 3\lambda_{RT}$ (i.e. three waves do grow on the disc surface) leads to the equation

$$\left(\frac{x}{r}\right)^4 = 12^2 \pi^2 \frac{1}{C_d We} \qquad (12)$$

And setting $x = 2r$ leads to the minimum value

$$We_{min} = \frac{144\pi^2}{16 C_d} = 52 \qquad (13)$$

for the onset of the so-called bag-and-stamen or umbrella breakup. To conclude this analysis, it can be said that this quite simple modeling leads to a proposed range for the bag breakup of [23-52] compatible with previous and present (only the largest peak is in this range) incompressible measurements.

## 5   *Turbulent cascading mechanism*

During the preceding events, numerous tiny droplets are produced as a result of the burst of the bag. Greatly different from the first two narrow peaks, their size distribution is very wide. This section will be devoted to the interpretation of this widespread distribution of fragments.

It is postulated thereafter that this widespread distribution is the image of the widespread distribution of vortices in turbulent flows, a phenomenon known as intermittencies. In [12] a detailed scenario of turbulent intermittencies is devised thank to a self-avoiding random vortex stretching mechanism which ultimately results in a log-stable distribution of vortices. In this work, in a very classic way, the size of the most common vortices is given by the Taylor micro scale and the size of the smallest vortices by the Kolmogorov scale. Stable distributions are defined by four parameters. These are proved to be $\alpha$ = 1.70 (theoretically) and 1.68 (experimentally) for the stability index, $\beta$ = -1 for the asymmetry parameter (both theoretically and experimentally, the resulting distributions are said to be totally skewed to the left), the scale parameter is given by

$$\sigma_{\ln\varepsilon}^{\alpha} = \ln\left(\frac{\lambda}{\eta}\right), \tag{14}$$

and the shift parameter $\delta_{\ln\varepsilon}$ is given by the average dissipation of turbulent kinetic energy per unit volume and is therefore related to the large scale of the flow.

It is in the present experiment quite difficult to estimate *a priori* the different scales of turbulence. Setting some reference values for the turbulent kinetic energy and the large scale of is required [32], let us try to do so:

$$\varepsilon \cong \frac{u'^3}{L_{int}} \cong \frac{2^3}{1350.10^{-6}} \cong 5900 \ m^2/s^3 \tag{15}$$

$$\lambda = \sqrt{20\nu\frac{k}{\varepsilon}} \cong 137 \mu m \text{ (Magnitude 2.13)} \tag{16}$$

$$\eta \cong \left(\frac{\nu^3}{\varepsilon}\right)^{1/4} \cong 3\mu m \text{ (Magnitude 0.48)} \tag{17}$$

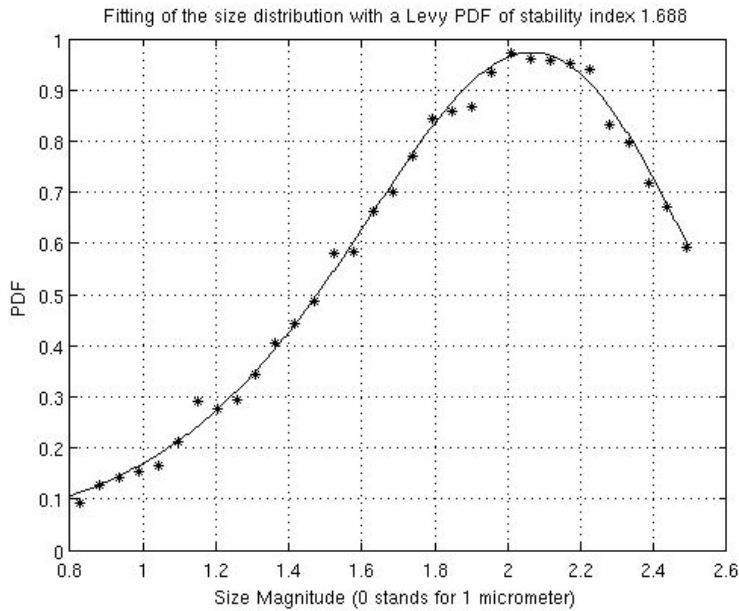

**FIG. 6: The bag-breakup wide peak is fitted with a log-stable distribution of stability parameter close to 1.68**

Note that the value of the turbulent fluctuating velocity inside the liquid phase has been identified with the RMS fluctuation of the droplets velocity and is therefore as likely to represent the fluctuation of the air velocity; anyway the value of turbulence dissipation rate $\varepsilon$ can thus be quite underestimated. However the resulting Taylor micro scale seems to be a good order of magnitude of the most common droplets. Yet some droplets whose sizes are located under the estimated Kolmogorov scale are also observed. While this could be related to the underestimation of the turbulence dissipation, this could also be related to some $2\pi$-phase ambiguity of the measuring device where big droplets can be interpreted as very small one (hopefully they are not very numerous). To circumvent this potential problem and to stay closer to the 1.6 decade dynamic, the bag number PDF has been fitted with a log-stable law (cf. [16] for the fitting procedure) in the magnitude range [0.8, 2.5]. The result is shown in figure 6. The value of the stability parameter is found to be 1.688 very close to the experimental value of 1.684 found in [12] for turbulent intermittencies (a fitting on the [0,2.5] magnitude range gives the value 1.70). Using (14) and values (15), (16), (17), the expected value of the scale parameter for the turbulent dissipation is found to be $\sigma_{ln\varepsilon}$ = 2.2. Present fitting of droplets PDF led to a scale parameter $\sigma_{lnd}$ equals to 1.1 i.e. about half the scale parameter of the turbulent dissipation.

$$\sigma_{\ln d} \approx \tfrac{1}{2} \sigma_{\ln \varepsilon} \tag{18}$$

It seems, at first, rather difficult to relate, theoretically, this new scale parameter to the previous turbulent intermittency scale parameter. It is well known that breakup of the bag leads to the formation of filaments which then turn into droplet [31], therefore volume of the droplet shall be related in a way to the size of these filaments. However the very nature of these filaments remains unknown .As they are seemingly coherent structures, a possible explanation could be that they are composed of vortex filaments resulting from a turbulent vortex cascading mechanism. Yet, this consideration can only be qualitative.

A possible scenario leading to more quantitative value, has been given by Hinze [33,15] when he devised a mechanism of droplets breakup by the turbulence of the carrier phase. It can be easily adapted to a situation where the turbulence inside the (carried) fluid leads to interface creation and to the formation of droplets. Let us equate the turbulent dynamic pressure at the surface, induced by either inner or outer movements, to the surface tension pressure of a cylindrical filament of diameter $d$ centered in $x$ (this leads to a turbulent breakup condition for a droplet of diameter $d$):

$$\tfrac{1}{2} \rho_L \left( u(x+d/2) - u(x) \right)^2 = \frac{\gamma}{d} \tag{19}$$

Then, using Kolmogorov 4/5$^{th}$ law, one gets:

$$d = \left( \frac{2\gamma}{\rho_L} \right)^{3/5} \left( \tfrac{5}{2} \right)^{2/5} \varepsilon_d^{-2/5}, \tag{20}$$

Therefore

$$\ln(d) = -\tfrac{2}{5} \ln(\varepsilon) + cst \tag{21}$$

and the following relation between scale parameters could be expected:

$$\sigma_{\ln d} = \tfrac{2}{5} \sigma_{\ln \varepsilon} \tag{22}$$

This leads to an expected value of 0.9 for $\sigma_d$ whereas 1.1 was measured. Agreement is not perfect and could be due to the harsh estimates of the different turbulent scales and to some uncertainty on the parameter estimation of the log-stable law. However the minus sign in (11) reverse the skewness of the stable distribution and the skewness parameter of the distributions of $\ln(d)$ and $\ln(\varepsilon)$ shall be equal but opposite. It is therefore not compatible with present measurements ($\beta_{lnd}$ = -1) and the log-stable turbulence model ($\beta_{ln\varepsilon}$ = -1). Moreover a simple computation of the order of magnitude in (10) leads to:

$$d = \left( \frac{2 \times 0.072}{1000} \right)^{3/5} \left( \tfrac{5}{2} \right)^{2/5} (5900)^{-2/5} \approx 221 \mu m \text{ (magnitude 2.3)} \tag{23}$$

This gives a good order of magnitude of the Sauter Mean Diameter of the bag but also indicates that smaller droplets shall be stable. Both these arguments indicate that inner turbulence in the mother droplet cannot be considered to be the mechanism by which the droplets resulting from the bag breakup do appear. Note that considering that turbulent movements of the air leads to the breakup of droplets(i.e. replacing $\rho_L$ by $\rho_G$ as Hinze originally did) leads to higher values in (23)

Therefore finding another model seems necessary. Liu and Reitz observed [31] that the breakup of the bag resulted in numerous tiny filaments which then reorganize into droplets. These ligaments can then be covered, according to P$^r$ Villermaux [20,34], by small balls (blobs) which add up their volume to make the final droplet. By making some hypotheses about so-called interaction layers, which can be (roughly) understood as a hypothesis of independence of the size of the covering balls whose radius are supposed to be exponentially distributed [35], this results in a self-convolution process leading to a gamma PDF (i.e. a product of a power law and a exponential). The conclusion is interesting but some hypothesis concerning these interaction layers seems questionable [34]. It is known that lognormal PDF can be replaced by gamma PDF in some turbulence modeling [36], but this make intermittency disappear as the tail of the distribution is much shorter. Unfortunately, since high-speed imaging leads to a limited range of droplets size, P$^r$ Villermaux and coworkers use less than a decade to comfort their model. Nevertheless, the physical insight contained in their work seems interesting and may be adapted to give another interpretation of present bag droplets PDF.

Let us suppose that the fundamentals blobs are of radii equal to the Kolmogorov scale and that the interaction layers are the results of the agitation by the turbulence of the flow. Since the surface energy needed to form the filament does not seem to come from the turbulent motion developing inside the water (cf. (23)), let us suppose that this energy directly come from the average kinetic of the fluid (the minimum droplet size in this case is $d = 2\gamma/\rho U^2 \cong$ 90nm much below what is observed; nevertheless this also means that such an energy source is available) and that the mixing of the spray with the air is the result of the natural expansion of the jet. Let us suppose that coherent structures developing inside the water are more able to resist to this mixing process but ultimately recess thanks to an agglomeration process. Then the size distribution $n(d,t)$ of droplets can given by Smoluchowski's equation [37]:

$$\frac{\partial n(d,t)}{\partial t} = \tfrac{1}{2}\int_0^d a(d-\xi,\xi) n(d-\xi,t) n(\xi,t) d\xi - n(d,t)\int_0^\infty a(d,\xi) n(\xi,t) d\xi \qquad (24)$$

where $a$ is the aggregation kernel. If $a$ is supposed constant this equation has an analytical solution [38] and the first order moments of the PDF are given by:

$$n(t) = m_0(t) = \frac{2n_0}{2+an_0 t} \qquad (25)$$

$$m_1(t) = d_0(n_0 - m_0(t)) \qquad (26)$$

So that the average length reads:

$$d = \frac{m_1(t)}{m_0(t)} = d_0 \frac{an_0 t}{2} \qquad (27)$$

$d_0$ is the initial value of the blobs diameter or filaments thickness, $n_0$ is their initial numeric density per unit volume, hence the product of these two quantities can be considered as the cumulated filament length per unit volume. Let us suppose that the aggregation speed is governed by the gradient of the fluctuating velocity i.e. $a \sim \partial u/\partial x \sim \sqrt{\varepsilon/\nu}$ i.e. is the inverse of Kolmogorov turbulent time $\tau_\eta$. Let us suppose that this aggregation characteristic time is much shorter than the overall aggregation time $t$ which will be supposed equal to the turbulent integral time $\tau_{int}$ i.e. aggregation can be supposed to happen inside the largest eddies.
Using standard approximation [32], the integral scale can be written

$$\tau_{int} = \frac{L_{int}}{k^{\frac{1}{2}}} \approx \frac{1350.10^{-6}}{6^{\frac{1}{2}}} = 551 \mu s \quad (28)$$

which is, as can be expected, much larger than the Kolmogorov time scale given by

$$\tau_\eta = \sqrt{\frac{\nu_G}{\varepsilon}} \approx \sqrt{\frac{1.5 10^{-5}}{5900}} \approx 50 \mu s \quad (29)$$

Using (27), (28) and (29), the droplet size can be written:

$$d = \frac{m_1(t)}{m_0(t)} \approx d_0 \frac{n_0}{2} \frac{\tau_{int}}{\tau_\eta} = \frac{l}{2} \frac{L_{int}}{k^{\frac{1}{2}}} \sqrt{\frac{\varepsilon}{\nu_G}} \propto \varepsilon^{\frac{1}{2}} \quad (30)$$

Therefore the expected scaling ½ between droplet size and turbulent energy dissipation (cf. Eq. 18) is recovered. Actually, the present modelling has been devised in order to recover this scaling. Actual test of this dependency is however quite difficult as it is hard to make sensible changes to the turbulence parameters without changing either the atomization device or the atomization regime. However the same dependency has been recovered for a very different atomization mechanism: a full cone pressure swirl nozzle [39], this therefore seems to indicate that this model has some relevance. Further tests will however be required to verify this dependency. To conclude with this section, let us pinpoint that the present model is very close to some micro mixing model developed in chemical engineering [40] where Kolmogorov time scale is the reference mixing time.

## 6    *Conclusion*

In this work, it has been shown that for some high flow-rate industrial spray, in the bag-breakup atomization mechanism, the drop PDF was composed of three peaks. The first two peaks are narrow and correspond respectively to the mother droplet peak and to the peak of daughter droplets created by burst of the basal ring, or rim of the bag. The ratio 2:1 between these two peaks can be adequately explained by a combination of Rayleigh-Taylor instability and Droplet Distortion and Breakup modeling. An interesting upshot of this model is that it accurately predicts the transition between the bag-breakup and the bag-and-stamen or umbrella breakup regime. The third peaks related to the bag breakup leads to a very wide range of fragments. The most common size of these fragments can be related to the Taylor scale or size of the most common turbulent eddies. These eddies possibly develop inside the droplet during the formation of the bag and its subsequent bursting into numerous filaments. The size PDF of the resulting droplets is very close to log-stable distribution used in turbulence intermittency modeling and a possible scenario is built in order to explain the values of the observed parameters and their relative dependency: the value of the stability index is found to be equal to 1.68 as in turbulence intermittency experiments (theory leads to a value close to 1.70), the value of the skewness parameter is set to -1 and the value of the scale parameter seems to be half the value of the scale parameter of the turbulence energy dissipation distribution. The value of the shift parameter is set by the large scale of the flow i.e. by the wavelength of the main instability mechanism, here Rayleigh-Taylor instability. Though the relationship between scale parameters has still to be investigated more thoroughly, let us pinpoint that one of the main advantage of the present modeling is that it does not introduce any new parameter. It is our hope that the physical insight contained in our model will help in the design of future nozzles.

## 7    *Acknowledgement*

The authors thank D$^r$. A. Delconte and D$^r$. A. Labergue for their help during data acquisition and P$^r$. B. Oesterlé and P$^r$ F. Lemoine for helpful discussions and comments.

# 8     *Appendix*

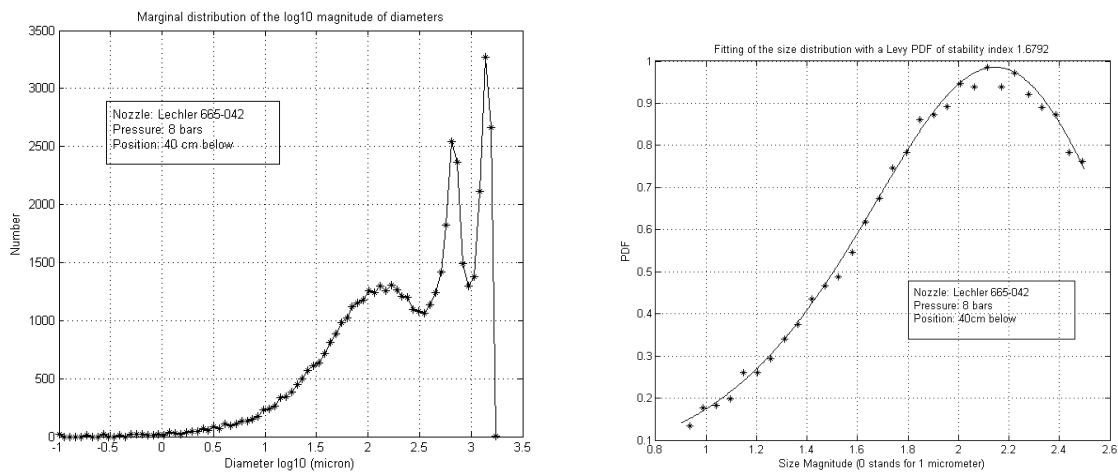

**FIG. 7: Results for the 665-042 Lechler Nozzle 40 cm below the exit/**